\documentclass[twocolumn,preprintnumbers,superscriptaddress,prb,floatfix]{revtex4}
\usepackage{amsmath}
\usepackage{amssymb}
\usepackage{color}
\usepackage{epsfig}
\usepackage{dcolumn}
\usepackage{bm}

\newcommand{\ud}[1]{#1^{\dag}}
\newcommand{\bra}[1]{\left\langle #1\right|}
\newcommand{\ket}[1]{\left| #1\right\rangle}

\newcommand{\Tr}{{\rm Tr}}

\begin{document}

\title{Optical spectra of a quantum dot in a microcavity in the nonlinear regime}
\author{E. del Valle}
\affiliation{Departamento de F\'isica Te\'orica de la Materia Condensada, Universidad Aut\'onoma de Madrid, Spain}
\affiliation{International Centre for Condensed Matter Physics, Universidade de Bras{\'i}lia, 70904-910, Bras{\'i}lia-DF, Brazil}
\author{F. P. Laussy}
\affiliation{Departamento de F\'isica Te\'orica de la Materia Condensada, Universidad Aut\'onoma de Madrid, Spain}
\affiliation{International Centre for Condensed Matter Physics, Universidade de Bras{\'i}lia, 70904-910, Bras{\'i}lia-DF, Brazil}
\author{F. M. Souza}
\affiliation{International Centre for Condensed Matter Physics, Universidade de Bras{\'i}lia, 70904-910, Bras{\'i}lia-DF, Brazil}
\author{I. A. Shelykh}
\affiliation{Faculty of Science, University of Iceland, Dunhaga 3, IS-107, Reykjavik, Iceland}
\affiliation{International Centre for Condensed Matter Physics, Universidade de Bras{\'i}lia, 70904-910, Bras{\'i}lia-DF, Brazil}
\affiliation{St. Petersburg State Polytechnical University, 195251, St Petersburg, Russia}


\begin{abstract}
  The optical emission spectrum of a quantum dot in strong coupling
  with the single mode of a microcavity is obtained in the nonlinear
  regime. We study how exciton-exciton interactions alter the emission
  spectrum of the system, bringing the linear Rabi doublet into a
  multiplet structure that is strongly dependent on the cavity-exciton
  energy detuning. We emphasise how nonlinearity can be used to
  evidence the genuine quantum nature of the coupling by producing
  satellites peaks of the Rabi doublet that originate from the
  quantized energy levels of the interactions.
 \end{abstract}

\volumeyear{year} \volumenumber{number} \issuenumber{number}
\eid{identifier}
\date[Date: ]{\today}
\maketitle

\section{Introduction}

Since the first experimental realization of atomic cavity quantum
electrodynamics (cQED),\cite{fs_haroche89a} a wealth of new results
have been obtained in the physics of strong coupling of a zero
dimensional quantum system to a single photon
mode.\cite{hood98a,boca04a,aoki06a,srinivasan07a,englund07a} The
problem is important not only because of the fundamental aspects
brought forward by the interaction of material systems with
photons,\cite{mabuchi02a} but also because of the potential
application of cQED to quantum information
processing.\cite{imamoglu99a,bennett00a} In semiconductors, strong
coupling is routinely achieved in planar microcavities---where
in-plane excitons couple to photons with matching momenta---since the
phenomenon was experimentally evidenced in 1992.\cite{weisbuch92a}
Strong coupling in 2D now provides the basis for investigating exotic
phases of matter in these structures such as Bose-Einstein
condensates\cite{kasprzak06a,balili07a,brennecke07a,colombe07a} or
superfluids.\cite{malpuech03b,carusotto04a,shelykh06a,fs_amo07a} One
system that has recently attracted particular attention is the
zero-dimensional analog that consists of a Quantum Dot (QD) coupled to
a single microcavity
mode.\cite{andreani99a,reithmaier04a,yoshie04a,peter05a,kaliteevski07a}
The material excitations in the QD are excitons, that is, bound
electron-hole pairs. Owing to their spatial confinement and energy
level discretization, they can be brought in strong coupling with the
single mode of a microcavity, such as that offered by a pillar (etched
planar cavity),\cite{reithmaier04a} the defect of a photonic
crystal\cite{yoshie04a} or the whispering gallery mode of a
microdisk,\cite{peter05a,kaliteevski07a} among others. In
Refs.~[\onlinecite{reithmaier04a,yoshie04a,peter05a}], such structures
have demonstrated the Rabi doublet in their optical spectra, which is
characteristic of the mode anticrossing that marks the overcome of
dissipation by the coherent exciton-photon interaction.

Steady progress is being made towards a better quantum strong-coupling
as well as towards its external control. Passive photonic crystals
with breathtaking values of~$Q=2.5\times10^6$ have now been
reported,\cite{noda07a} limited only by structural imperfections.  As
the fabrication process will improve, a gain of another order of
magnitude in the quality factor is allowed by the available
theoretical designs.  As for the active element, which inclusion
lowers the quality factor, prowess have been achieved in the
engineering of the heterostructure allowing to deterministically
position a QD inside a photonic crystal to within 25nm accuracy, and
thus place the dot at a maxima of the light intensity, along with an
etching technique of the holes of the photonic crystal to match
spectrally the QD and cavity mode emission.\cite{badolato05a}
Micropillars etched out of Bragg mirrors allow a fine control of the
radius and even of the shape of the cross-section, that can be optimised
to reach high values of the quality factor, with figures
of~$1.65\times10^5$ reported for radius size
of~$4\mu$m.\cite{reitzenstein07a} One appealing feature of these
structures is their ease of access both for the excitation and the
emission, the latter being perpendicular to the sample surface. The
mode volume~$V$ in these structures can be further reduced while
maintening the quality factor in one polarization by using elliptical
pillars, and record values of~$Q/\sqrt{V}$ can be achieved in this
way.\cite{whittaker07a} At the same time, the density of
self-assembled QDs in the active medium has been successfully reduced
over the years, with figures of~$10^{10}$cm$^{-2}$, and the
possibility to grow large dots (with lens shape of~$\approx30$nm) so
as to provide a large oscillator strength.  With microdisk resonators,
the whispering gallery modes of a thin disk supported by a column
provide the high~$Q$ modes of the cavity.

These milestones open the way to new research investigating the
``true'' quantum nature of the exciton-photon
coupling.\cite{khitrova06a} Although the system exhibits strong
coupling, it is not known in which quantum state it is actually
realized. To be useful for quantum information, the system must deal
with few quanta of excitation and display strong alterations of its
characteristic for small changes in particle number populations.  If
the system scales linearly with the number of particles, it is
essentially classical,\cite{zhu90a} and adding or removing a single
particle will not change its behaviour.  Therefore, probing the
nonlinear response is desirable.  Recently, in an attempt to validate
the single-particle character of the Rabi doublet in a QD-microcavity
system, an unexpected additional peak was observed that was attributed
to renormalisation of the exciton energy by charging centres close to
the QD.\cite{hennessy07a} The observed antibunching allowed the
authors of Ref.~[\onlinecite{hennessy07a}] to confirm the quantum
character of this coupling, but much is left to understand of the
nonlinear effects and the nature of the additional peaks, observed or
predicted.

In this text, we compute the spectra of emission for a pumping high
enough to enter the nonlinear regime. With QDs in microcavities, two
types of strong nonlinearities are expected,\cite{laussy06b} both
associated to the active material, i.e., the excitons. The first one
comes from Coulomb repulsion of the charged particles, and is the one
investigated here, in the case where it is comparable to the coupling
strength. The other comes from Pauli exclusion, that arises from the
fermionic character of the underlying particles.\cite{combescot04a} In
Refs.~[\onlinecite{laussy06b,laussy05c}], the effect of Pauli
exclusion alone was investigated (Coulomb interactions were added in
an as yet unpublished work.\cite{fs_glazov}) There it was shown that
Pauli exclusion resulted in rich multiplet structures depending on the
size of the dot, with a Mollow triplet-like structure for small dot
with strong Pauli repulsion, and a Rabi doublet for large dots. In the
present work, we show that even for bosonic excitons, a multiplet
structure arises in the emission spectrum when intra-dot
exciton-exciton interaction is accounted for.  For the large dots that
the model aims to describe, Pauli exclusion can be taken into account
phenomenologically by a phase-space filling effect that screens the
exciton-photon
interaction.\cite{hanamura70a,schmittrink85a,imamoglu98a} The latter
results in loss of strong-coupling and we therefore focus on the
intermediate regime where such renormalization of the coupling
strength can be neglected. We also neglect the spin-degree of freedom,
in particular the sign-dependent interaction between same and
opposite-spins excitons, respectively.\cite{kavokin_book07a} This
allows us to focus on nonlinear deviations and to neglect more
complicated correlations effects of the multi-excitons complexes such
as formations of bound pairs or molecules that would give rise to
bipolaritons.\cite{ivanov98a,ivanov04a,gotoh05a} Experimentally, this
could be realized by using a circularly polarised pump. Strong
qualitative changes are still observed, with a basic structure that
consists of a blueshifted peak on top of a distorted Rabi doublet.

The paper is organised as follow.  In Sec.~\ref{secI}, we introduce
the Hamiltonian of the system describing the coupling between cavity
photons and interacting QD excitons. From this Hamiltonian, we extract
the energy level structure. This already contains the essential
information to understand qualitatively the spectra that we compute
through more refined methods in the next section. In Sec.~\ref{secMF},
we use the density matrix formalism and solve the master equation in
the steady state to compute the spectra using the quantum regression
theorem. This method, which suffers in principle no limitation or
approximation (beyond assuming Markovian dynamics), can only be used
in practice for the low pumping regime, as the computational cost goes
like~$N^8$ with the method that we use here, where~$N$ is the
truncation order (the maximum number of photons accounted for in the
model). We also propose an alternative formalism for the calculation
of the emission spectra based on the Keldysh Green function technique
and compare the results obtained by the two methods. In
Sec.~\ref{conclusions}, we discuss the overall results and
conclude. The Appendix contains the mathematical details of the
calculation of the emission spectra using the Keldysh Green function
technique.

\section{Model Hamiltonian}
\label{secI}

We describe light-matter interaction in a large QD including
exciton-exciton interactions with the following Hamiltonian:
\begin{multline}
  \label{eq:H}
  H= \sum_{j=1,2} \epsilon_j c_j^\dagger c_j {}+V_\mathrm{R}
  (c_1^\dagger c_2 + c_2^\dagger c_1) +\frac{U}{2} c_2^\dagger
  c_2^\dagger c_2 c_2.
\end{multline}

The operator $c_j$ ($c_j^\dagger$) is the annihilation (creation)
operator for the cavity photon ($j=1$) and QD exciton ($j=2$). They
all follow Bose algebra. The parameter $V_\mathrm{R}$ is the coupling
strength and $U$ accounts for the exciton-exciton interaction. In all
the following, $V_\mathrm{R}$ is taken as the unit. An important
parameter of the system is the detuning
$\Delta=\epsilon_1-\epsilon_2$.  The reference energy is that of the
exciton, that is set to zero, $\epsilon_2=0$.  In eqn~(\ref{eq:H}), we
neglected the polarization degree of freedom,\cite{shelykh05a} thus
addressing the case of a symmetric QD under normal excitation, or
excited by a circularly polarised light.

To obtain an intuitive picture on the optical spectrum of such a
system, we first study the allowed transitions between the possible
energy levels. That is, we first obtain the eigenstates of the system
with $n$ excitations by diagonalizing the Hamiltonian~(\ref{eq:H}).
Note that the total number of excitations is conserved. In what
follows, we refer to the set of eigenstates of the total particle
number~$\ud{c_1}c_1+\ud{c_2}c_2$ with eigenvalue~$n$ as the $n$th
\emph{manifold} of the system.

\begin{figure}[htbp]
\begin{center}
  \epsfig{width=0.45\textwidth,file=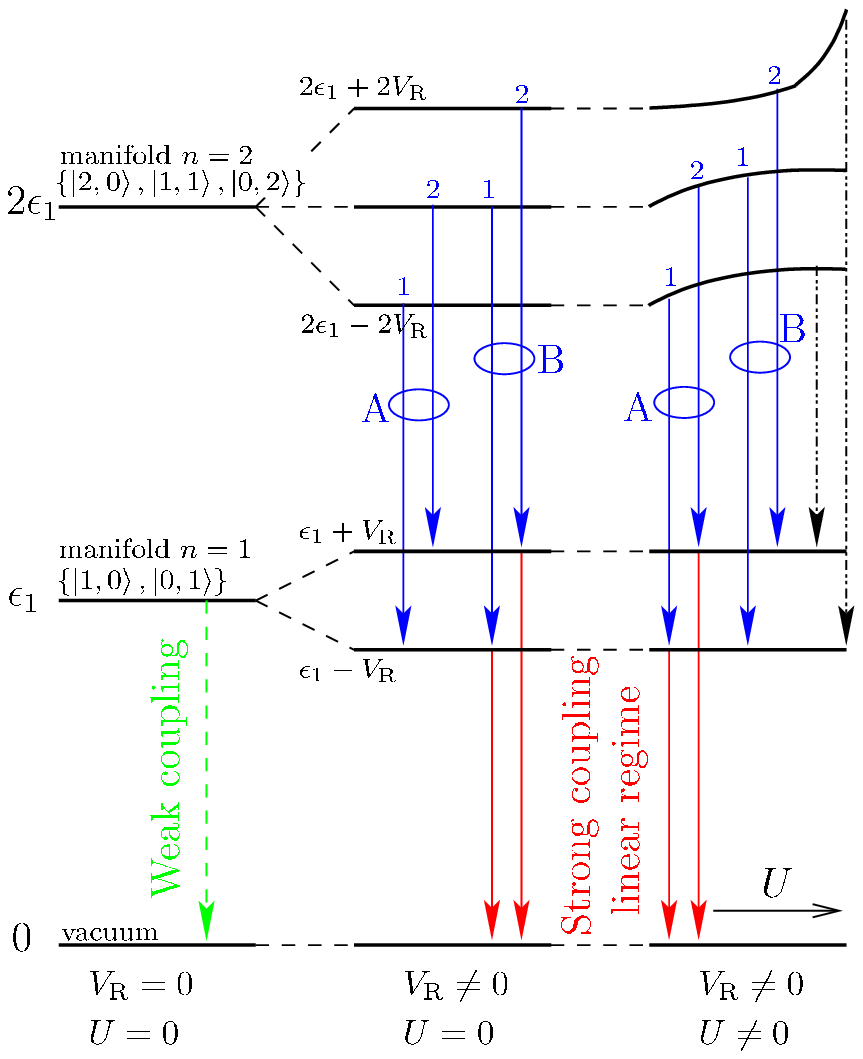}
\end{center}
\caption{(Colour online) Energy levels of the eigenstates of
  eqn~(\ref{eq:H}) up to the second manifold (two excitations), all at
  zero detuning ($\Delta = 0$), left panel for weak (or no-) coupling
  ($V_\mathrm{R} = 0$), central and right panel in strong-coupling,
  with right panel also including interactions~$U$ varying on the~$x$
  axis.  The transitions between levels account for the spectral
  features. Red lines correspond to the vacuum-field Rabi doublet,
  that turns into a single Green line in weak coupling regime. Blue
  lines superimpose to the Rabi doublet when higher manifolds are
  probed.  Without interactions, $U = 0$, these transitions do not
  appear in the spectra.  Transitions from $n=2$ to $n=1$ in presence
  of interactions are plotted in Fig.~\ref{figure3} as a function of
  the detuning and $U$.  New qualitative features appear thanks to the
  interactions. Black dashed lines are new transitions previously
  forbidden, although they remain weak.}
\label{figure1}
\end{figure}

\begin{figure}[htbp]
\begin{center}
  \epsfig{width=0.4\textwidth,file=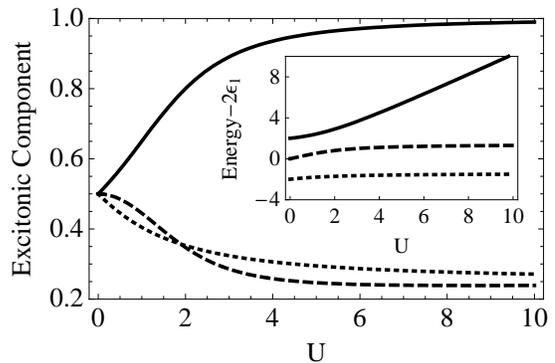}
\end{center}
\caption{Excitonic component at resonance of the eigenvectors
  corresponding to each energy level of the manifold $n=2$ [see
  Fig.~(\ref{figure1})] as a function of the interaction strenght $U$.
  Varying the detuning also changes the character of the lines. In
  inset is plotted the exact energies of the eigenstates as a function
  of~$U$, that are sketched in the right-upper part of
  Fig.~(\ref{figure1}).}
\label{figure2}
\end{figure}

The structure of energy levels at resonance is sketched in
Fig.~(\ref{figure1}) up to a maximum of two excitations (i.e., up to
the second manifold) for three different cases. First, the bare levels
corresponding to non-interacting and uncoupled (or weakly-coupled)
modes ($V_\mathrm{R}=0$, $U=0$). Second, the eigenenergies arising
from the coupling ($V_\mathrm{R}\ne0$, $U=0$), and finally, the
blueshifted lines that result from including the interactions
($V_\mathrm{R}\ne0$, $U\ne0$). All the levels but those in the
manifold $n=1$, involving only one particle, change with the
interactions because of the excitonic part of their corresponding
eigenvectors. In order to keep track of the excitonic character of
each level, in Fig.~(\ref{figure2}), we plot the excitonic component
of the eigenvectors of manifold $n=2$ and their corresponding
eigenenergies as a function of the interaction~$U$. We can see that,
starting from a situation completely symmetric between the photonic
and excitonic fractions, the higher level gets more and more
excitonic-like with $U$ and blueshifts strongly. The other two energy
levels are only slightly affected, as follows from their more photonic
character.  This characterisation of the levels, which also depends on
the detuning, plays an important role when identifying the spectral
lines, as we show in the following sections.

The energy levels of Fig.~(\ref{figure1}) are broadened by the finite
lifetime and by the incoherent pumping mechanism. This will be taken
into account accurately in Section~\ref{secMF}. At this stage, an
imaginary part is added to the energies of the bare states of the
Hamiltonian~(\ref{eq:H}). This is justified a posteriori by comparison
with an exact numerical method.

Once computed the new complex eigenvectors and eigenenergies, we
obtain the amplitudes of probabilities of loosing an excitation from a
given manifold to the neighbouring one counting one excitation less.
We consider the amplitude for the processes of annihilation of a
photon, that is, the transitions corresponding to the normal mode
emission. If the coupling is strong enough, the optical spectrum
$S_\mathrm{ph}(\omega)$ can be approximated as a sum of Lorentzians
corresponding to each allowed transition.\cite{laussy_arxiv} The
emission lines are finally broadened with the sum of the imaginary
parts of the eigenstates involved in the transition and weighted by
its probability.

\begin{figure}[htbp]
\begin{center}
  \epsfig{width=0.4\textwidth,file=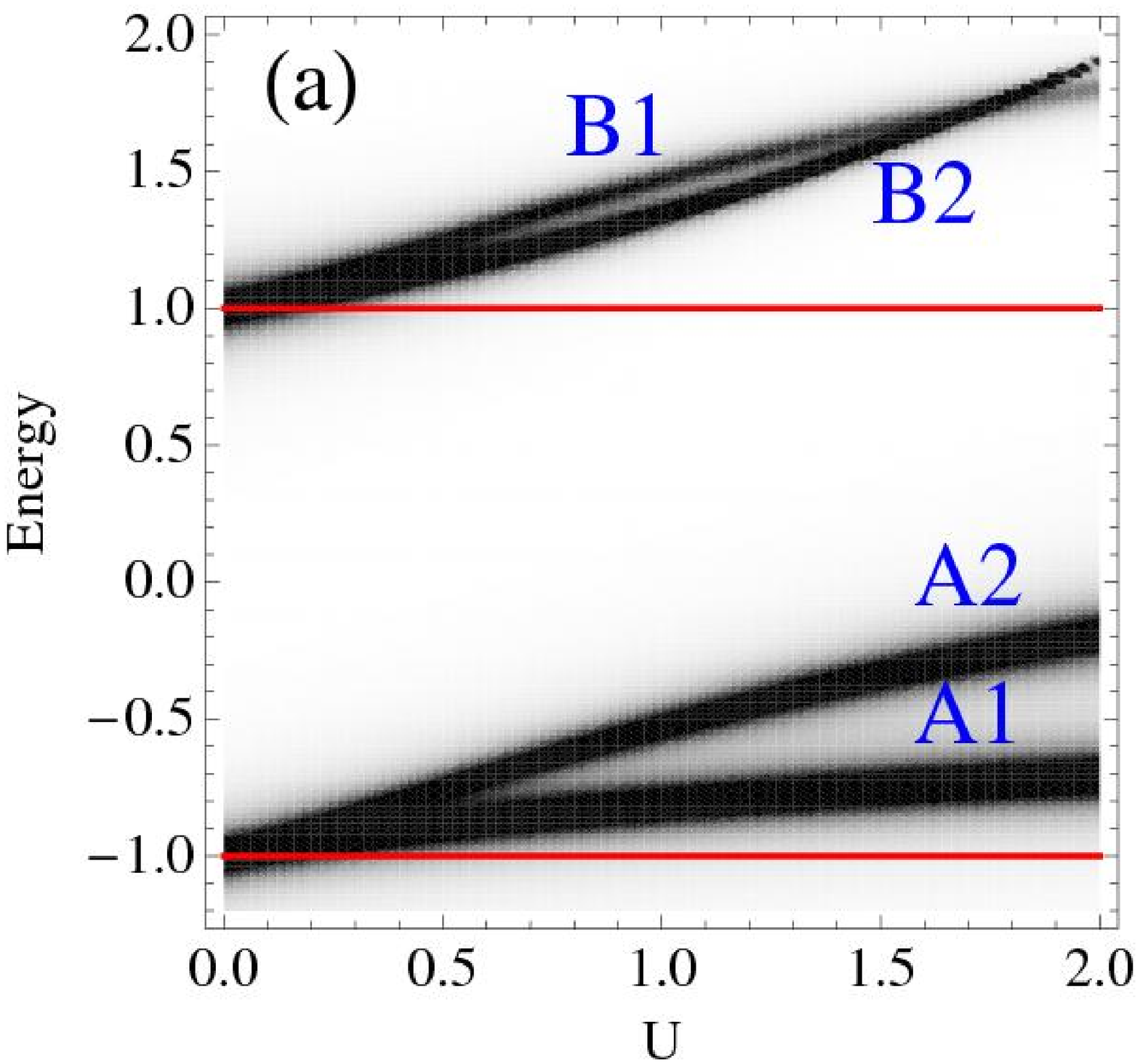}
  \epsfig{width=0.4\textwidth,file=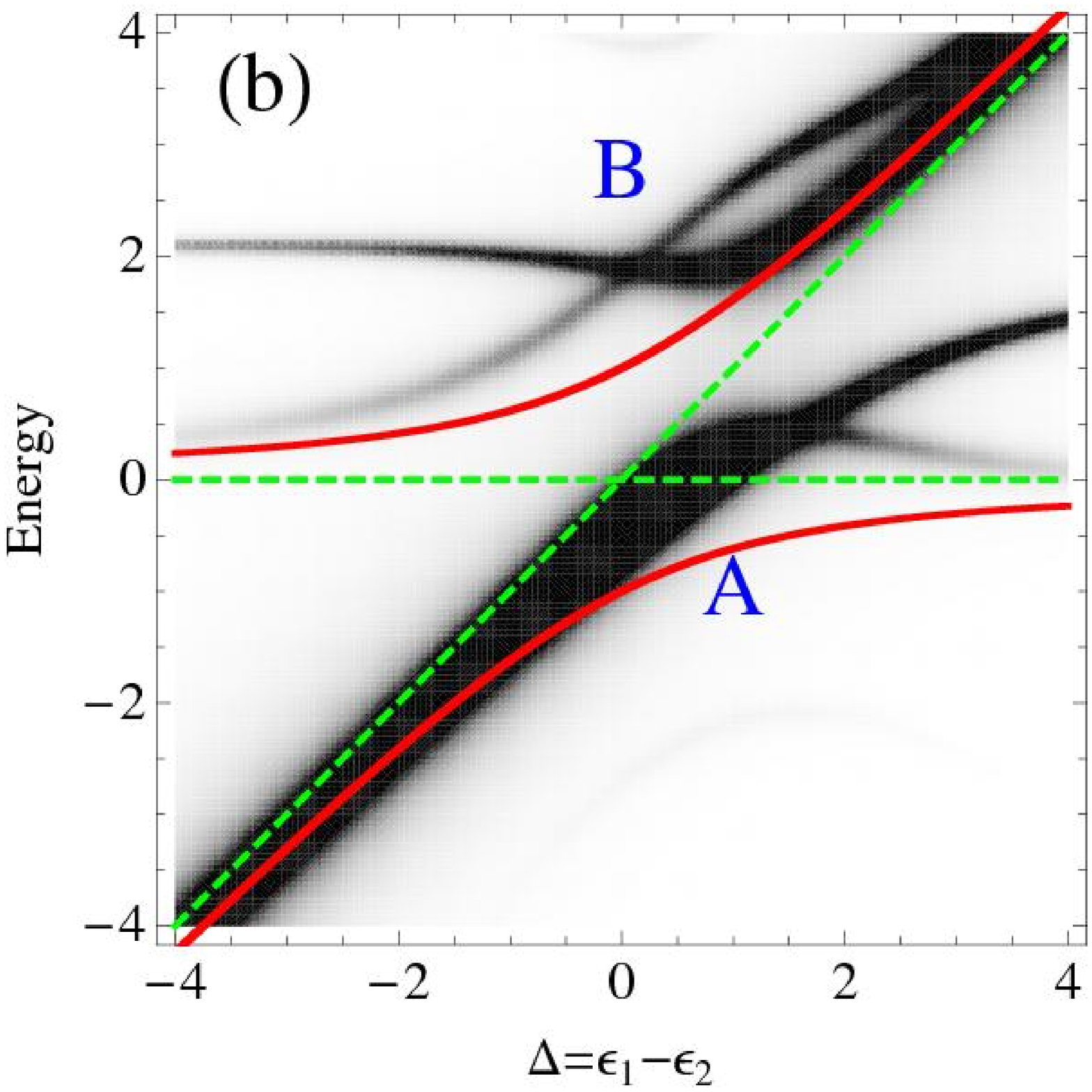}
\end{center}
\caption{(Colour online) Cavity emission spectra from second to first
  manifold transitions: (a) Resonant case as a funtion of interactions
  $U$ ($y$-axis). (b) Case of fixed interactions ($U=2$) as a funtion
  of detuning ($\Delta=\epsilon_1-\epsilon_2$, in $x$-axis). The
  non-interacting case~$U=0$, where only Rabi doublet arises, is also
  shown (red superimposed lines) as well as the bare cavity and
  excitonic lines (dashed green lines). Lines are labelled in blue
  corresponding to the transitions of Fig.~(\ref{figure1}).
  Parameters in both plots are: $\epsilon_2=0$, $\Gamma_1=0.1$,
  $\Gamma_2=0.01$, all in units of $V_R$.}
\label{figure3}
\end{figure}

These results are plotted for transitions between manifolds with 2 and
1 excitation as a function of the nonlinearity strength~$U$ on
Fig.~(\ref{figure3})a and as a function of the detuning on
Fig.~(\ref{figure3})b. In the latter case, the exciton bare energy is
kept constant, equal to zero, while the cavity mode is brought in or
out of resonance with the exciton.  This relative detuning between the
bare exciton and photon is tunable experimentally through a variety of
techniques that include changing temperature or applying a magnetic
field (shifting the energy of the dot with negligible perturbation on
the cavity) to growing a thin film (shifting the cavity mode energy
without affecting the dot). As this is a parameter easy to tune
experimentally, that brings many and specific changes in the system, a
density plot of photoluminescence emission with detuning of the type
of Fig.~\ref{figure3} is the kind of results we are aiming for.  On
such a plot, one can trace the intensity, width and location (that
include the number) of spectral lines for smooth changes in the
detuning.

Peaks appearing in Fig.~(\ref{figure3})a correspond to the transitions
plotted in Fig.~(\ref{figure1}).  They are labelled in blue (colour
online): lower lines are the transitions ``A'' and upper lines the
transitions~``B''. Comparing with the linear Rabi doublet, which is
superimposed in red, we observe the aforementioned blueshift of both
groups of lines. It is more important for the exciton-like mode
(especially line B-2 at resonance when~$U\gg1$) while the photon-like
mode has a better resolved fine-structure splitting. At various
detunings (Fig.  (\ref{figure3})b), complicated structures are found
with crossing or anticrossing of the lines, as shown on the figure.
Lines with the same bare-excitation (photon or exciton) character
cross, whereas lines of a different character exhibit anticrossing. At
large detunings, the bare photon and exciton modes (in green) are
recovered but with an additional blueshifted bare exciton line.

Satellite peaks arise at very low and high energies from transitions
that are forbidden in the linear regime. They enter the dynamics
through nonlinear channels opened by the interactions.  The dashed
arrows in the right panel of Fig.~(\ref{figure1}) represent these two
transitions, with two excitons as the initial state that release one
excitation and leave one photon as the final state. They appear dimly
in Fig.~(\ref{figure3}).

Thus, the manifold method allows an understanding of the composition
of the optical spectra, as illustrated on Fig.~(\ref{figure3}) where
the spectral lines have been labelled according to their corresponding
transitions on Fig.~(\ref{figure1}).  From this overall picture, the
excitonic fraction is clearly associated to the blueshift. Which
transitions enter the dynamical picture, and to what extent, is
investigated by the exact dynamical treatment that we pursue in terms
of the density matrix formalism.

In the next section, to describe more realistically the experimental
situation, we supplement eqn~(\ref{eq:H}) with an external excitation
and a decay mechanism for the QD exciton and cavity mode. This is done
by supplying a reservoir of external photons coupled to the cavity
mode (to account for the finite lifetime of cavity photons) and
another one to the exciton mode (to account for the exciton decay into
photonic modes other than the cavity). Both pumping and decay
compensate and the system reaches a steady state.

\section{Calculation of the emission spectra}\label{secMF}

The coupling of the system to the external world results in a
leakage of photons and of QD excitations, that imply that even when
the initial state of the system is well known (pure state), the
system evolves into a mixture of states where only probabilistic
information is available. A suitable treatment of such a dynamics
involves the density matrix operator~$\rho$. Its temporal evolution
is given by a master equation of the type:
\begin{equation}
  \label{eq:WedOct10211627UTC2007}
  \frac{d\rho}{dt}=i[\rho,H]+\frac{\chi}2\mathcal{L}_O\rho
\end{equation}
with~$H$ the Hamiltonian dynamics already introduced in
eqn~(\ref{eq:H}), and with the incoherent contributions in the form of
Lindblad terms~$\mathcal{L}_O$, defined as
\begin{equation}
  \label{eq:WedOct10211717UTC2007}
  \mathcal{L}_O\rho=2O\rho\ud{O}-\ud{O}O\rho-\rho\ud{O}O
\end{equation}
that corresponds to the general de-excitation operator~$O$ and its
effective decaying rate~$\chi$. Explicitly, the escape of the cavity
photons is accounted for by a Lindblad term~$\mathcal{L}_{c_1}\rho$,
with a rate $\chi=\Gamma_1$.  This parameter is inversely proportional
to the cavity quality factor~$Q$:
\begin{equation}
  \label{eq:WedOct10211845UTC2007}
  \Gamma_1=\epsilon_1/Q\,.
\end{equation}

The spontaneous decay of the QD excited states into any other mode
than the one of the cavity, as well as the non-radiative decay, are
taken into account with the term~$\mathcal{L}_{c_2}$ and its
associated rate~$\Gamma_2$.  Such a rate is typically much smaller
than the cavity emission rate, $\Gamma_1$.  Finally, the cavity or the
excitons can be pumped with a continuous incoherent pumping. The
associated Lindblad terms now involve the creation operators. The
master equation in our system therefore reads:
\begin{equation}
  \label{eq:WedOct10212018UTC2007}
  \frac{d\rho}{dt}=i[\rho,H]+\frac{\Gamma_1}2\mathcal{L}_{c_1}\rho+\frac{\Gamma_2}2\mathcal{L}_{c_2}\rho+\frac{P_1}2\mathcal{L}_{\ud{c_1}}\rho+\frac{P_2}2\mathcal{L}_{\ud{c_2}}\rho\,.
\end{equation}
We have included two possible kinds of incoherent pumping, namely, the
cavity pumping at rate~$P_1$ and the electronic pumping (via the
exciton) at rate~$P_2$. Experimentally, the most common practise is
electronic pumping, as the photoluminescence from the cavity is
usually observed and pumping it would hinder the measurement.  $P_1$
could be taken into account to describe an effective cavity pumping
due to, for instance, other dots that are weakly coupled to the cavity
modes and populate the cavity with photons~\cite{laussy_arxiv}. In the
case of direct cavity pumping, it would be more relevant to observe
the exciton emission for photoluminescence, or to turn to coherent
excitation and probe the transmission, reflexion and/or absorption of
the cavity.  Here we compute the cavity photoluminescence for the two
kinds of pumping each considered on their own, for the purpose of
comparison, keeping in mind that the more experimentally relevant case
is that of electronic pumping.

We first obtain the steady state $\rho^{(\mathrm{ss})}$ of the system,
by solving $d\rho/dt=0$.\cite{delvalle07b} From the knowledge of this
density matrix we can compute the mean value in the steady state of a
general operator~$\Omega$ as
$\langle\Omega\rangle=\Tr\{\rho^{(\mathrm{ss})}\Omega\}$. More
interestingly, the optical spectra of the system can also be obtained
from~$\rho^{(\mathrm{ss})}$ and the master
equation~(\ref{eq:WedOct10212018UTC2007}) thanks to the quantum
regression theorem.\cite{fs_eberly87,fs_molmer96a} We write the steady
state equation in the matrix form:
\begin{equation}
  \label{eq:WedOct10212313UTC2007}
  0=\frac{d\rho_{\alpha\beta}}{dt}=\sum_{\alpha'\beta'}M_{\alpha\beta,\alpha'\beta'}\rho_{\alpha'\beta'}
\end{equation}
where the labels~$\alpha$ and~$\beta$ index the whole Hilbert space,
namely, in our case of two oscillators, $\alpha=\{n,m\}$ and
$\beta=\{p,q\}$. As a result, $M_{\alpha\beta,\alpha'\beta'}$ is a
$N^4\times N^4$ matrix where~$N$ is the truncation of each
oscillator's Hilbert space. In the computations, we have checked that
the results were independent of this truncation once it is taken large
enough. The emission spectrum in the steady state, defined as
\begin{equation}
\label{Spec}
S_{\mathrm{ph}}(\omega)=\frac{1}{\pi}\mathrm{Re}\int_0^{\infty} e^{i \omega \tau} \langle c_1^\dagger(0) c_1(\tau)\rangle d\tau,
\end{equation}
can then be obtained from Lax's hypothesis that the two-time
correlators follow the same regression equations than the average:
\begin{equation}
  \label{eq:WedOct10212630UTC2007}
  S_{AB}(\omega) = \frac1{\pi}\mathrm{Re}\big(-\sum_{\alpha'\beta'}(M+i\omega \mathbb{I})^{-1}_{\alpha\beta,\alpha'\beta'}\rho^{(\mathrm{ss})}_{A;\alpha'\beta'}B_{\beta\alpha}\big)
\end{equation}
where~$A$ and~$B$ are the creation and destruction operators of the
transitions and $\rho^{(\mathrm{ss})}_{A;\alpha'\beta'}$ is defined as
\begin{equation}
  \label{eq:WedOct10212852UTC2007}
  \rho^{(\mathrm{ss})}_{A;\alpha'\beta'}=\sum_{i=1}\rho^{(\mathrm{ss})}_{\alpha'i}\bra{i}A\ket{\beta'}\,,
\end{equation}
that is, in the case of cavity (normal) emission, where~$A=\ud{c_1}$
and $B=c_1$:
\begin{align}
  \label{eq:WedOct10213017UTC2007}
  \rho^{(\mathrm{ss})}_{A,n,m;p,q}&=\sqrt{p+1}\rho^{(\mathrm{ss})}_{n,m;p+1,q}\,,\\
  B_{n,m;p,q}&=\sqrt{p}\delta_{m,q}\delta_{n,p-1}\,.
\end{align}

The drawback of this method is the high computational cost involved in
solving eqn~(\ref{eq:WedOct10212313UTC2007}) to obtain~$M$ and to
subsequently invert it in eqn~(\ref{eq:WedOct10212630UTC2007}).
Therefore, we present here the spectra for the low pump regime. This
is not a serious limitation for the time being, however, as we are
primarily interested in small deviations from the linear regime and to
probe the first ladders of the quantized energy levels. We use pumping
rates of~$0.01$, yielding average number of excitations of the
order~$\langle\ud{c_i}c_i\rangle\approx0.1$ with probability to have
two excitons of the order of $0.01$. For these figures, a truncation
at the fourth manifold ($N=3$) is enough to ensure convergence of the
results. The other parameters are fixed to the following values,
motivated by experiments: $V_R = 1$ provides the unit (experimental
figures are of the order of tens of $\mu$eV), $\Gamma_1=0.1$,
$\Gamma_2=0.01$ and $U=2$.

\begin{figure}[htbp]
  \centering
  \epsfig{file=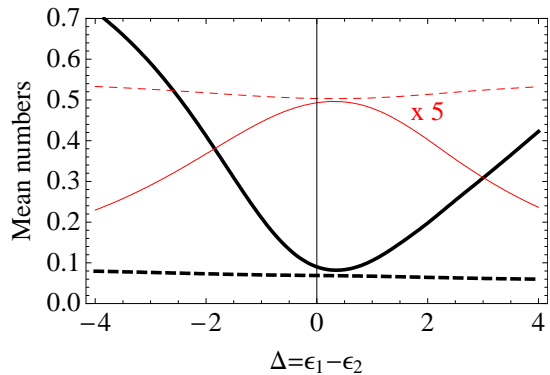, width=0.4\textwidth}
  \caption{(Colour online) Mean number of photons (dashed line) and
    excitons (solid line) as a function of the detuning between cavity
    and excitonic modes ($\Delta=\epsilon_1-\epsilon_2$), for the case
    of cavity pumping ($P_1=P$, $P_2=0$) in thin red ($\times5$) and
    electronic pumping ($P_1=0$, $P_2=P$) in thick black. Regardless
    of the detuning, an approximately constant and equal population is
    obtained for the cavity intensity, due to the balance between the
    effective coupling-strenght and the exciton-population.  An
    asymmetry is observed with detuning due to the interactions that
    bring the exciton closer or further to resonance with the cavity
    mode.  Parameters: $\epsilon_2=0$, $U=2$, $\Gamma_1=0.1$,
    $\Gamma_2=0.01$, $P=0.01$.}
\label{figure4}
\end{figure}
\begin{figure}[tbp]
\begin{center}
  \epsfig{file=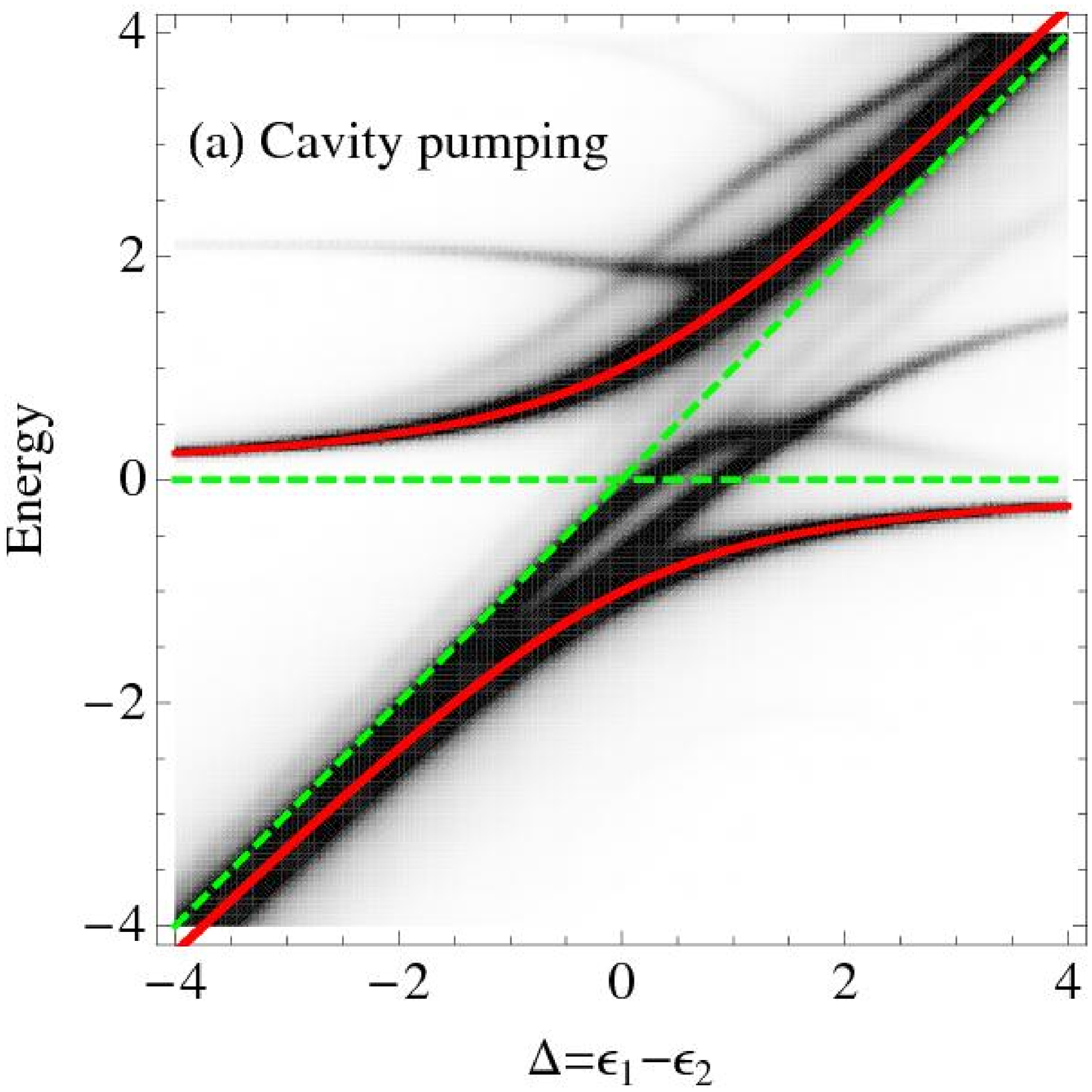, width=0.4\textwidth}
  \epsfig{file=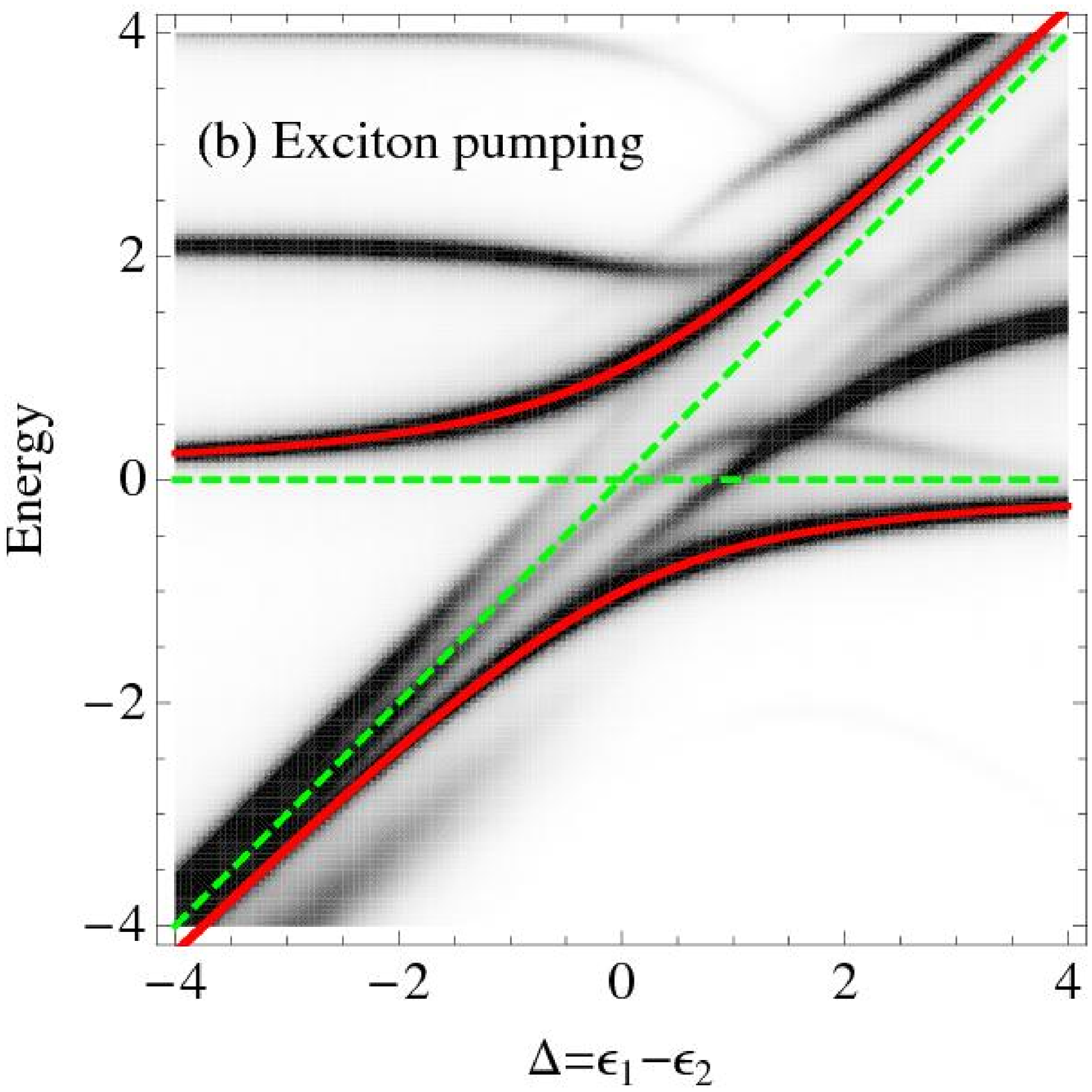, width=0.4\textwidth}
\end{center}
\caption{Cavity emission spectrum as a function of detuning $\Delta$.
  The $U=0$ case for photons, which corresponds to the Rabi splitting,
  is also shown in red solid lines as well as the uncoupled
  ($V_\mathrm{R}=0$, $U=0$) cavity and excitonic lines in green dashed
  lines. As the mean number of excitons is very low (smaller than 1)
  the probability of having more than two excitons is very low, only
  the second to first and first to zero manifold transitions appear.
  For high positive detuning, Coulomb interactions generate an
  additional peak close to $\epsilon_2+U(n_2-1)$. Note that peaks
  originated from transitions between manifolds with $n>2$ appear with
  a very small intensity, due to the non-zero probability to have
  three excitons in the system. All energies are in units of
  $V_\mathrm{R}$.  Parameters: $\epsilon_2=0$, $U=2$, $\Gamma_1=0.1$,
  $\Gamma_2=0.01$, $P=0.01$.}
\label{figure5}
\end{figure}

\begin{figure}[tbp]
\begin{center}
  \epsfig{file=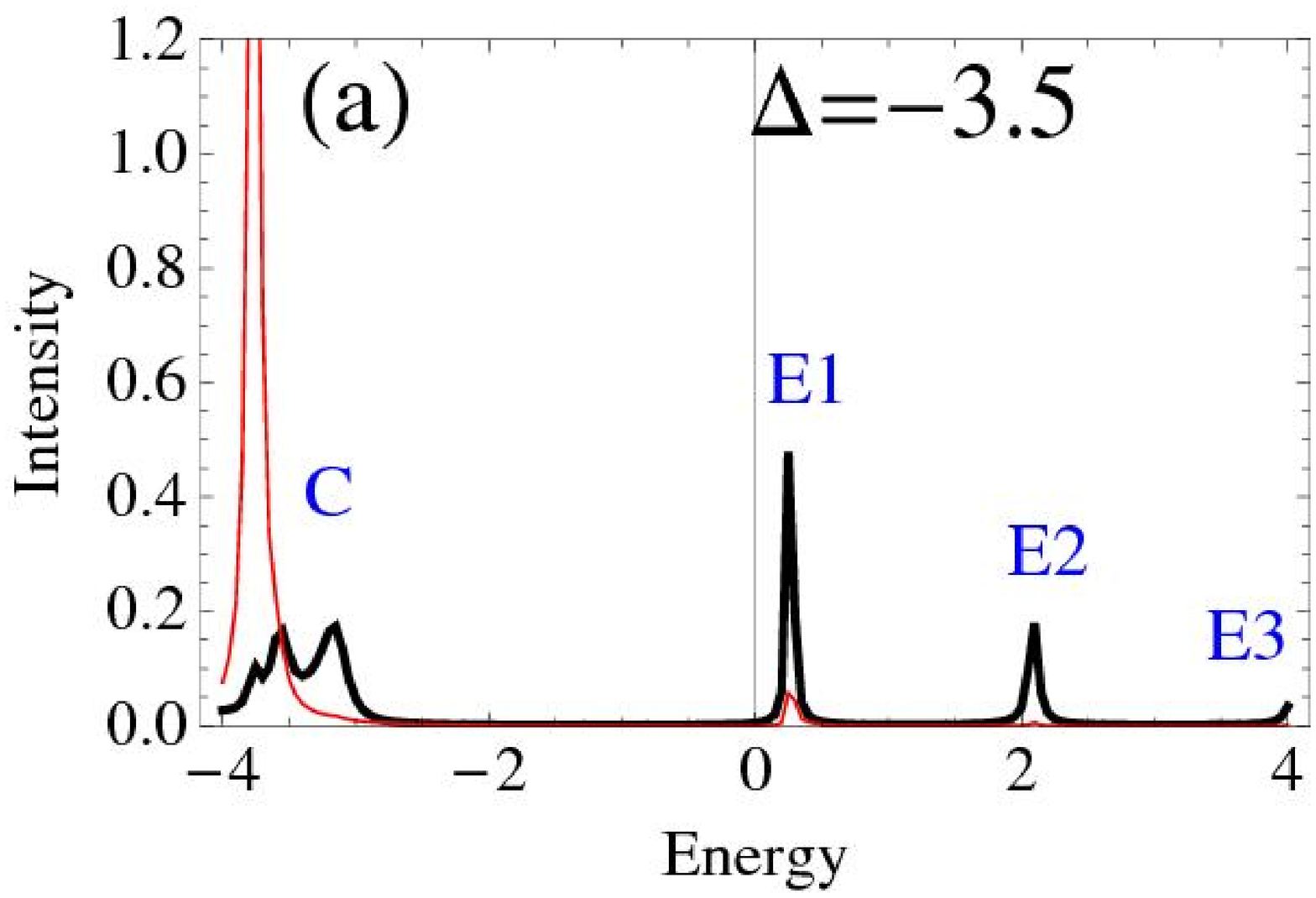, width=0.3\textwidth}
  \epsfig{file=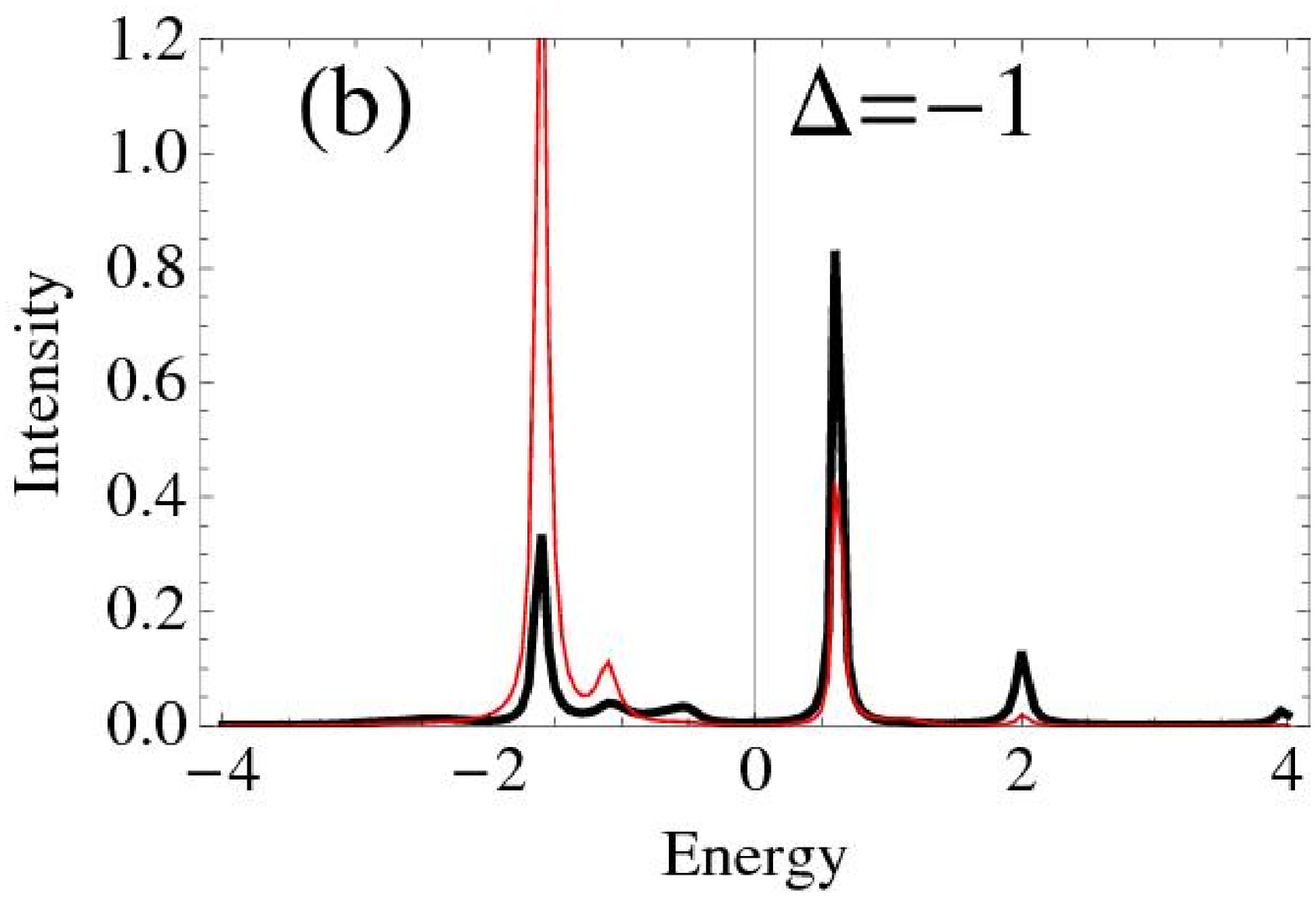, width=0.3\textwidth}
  \epsfig{file=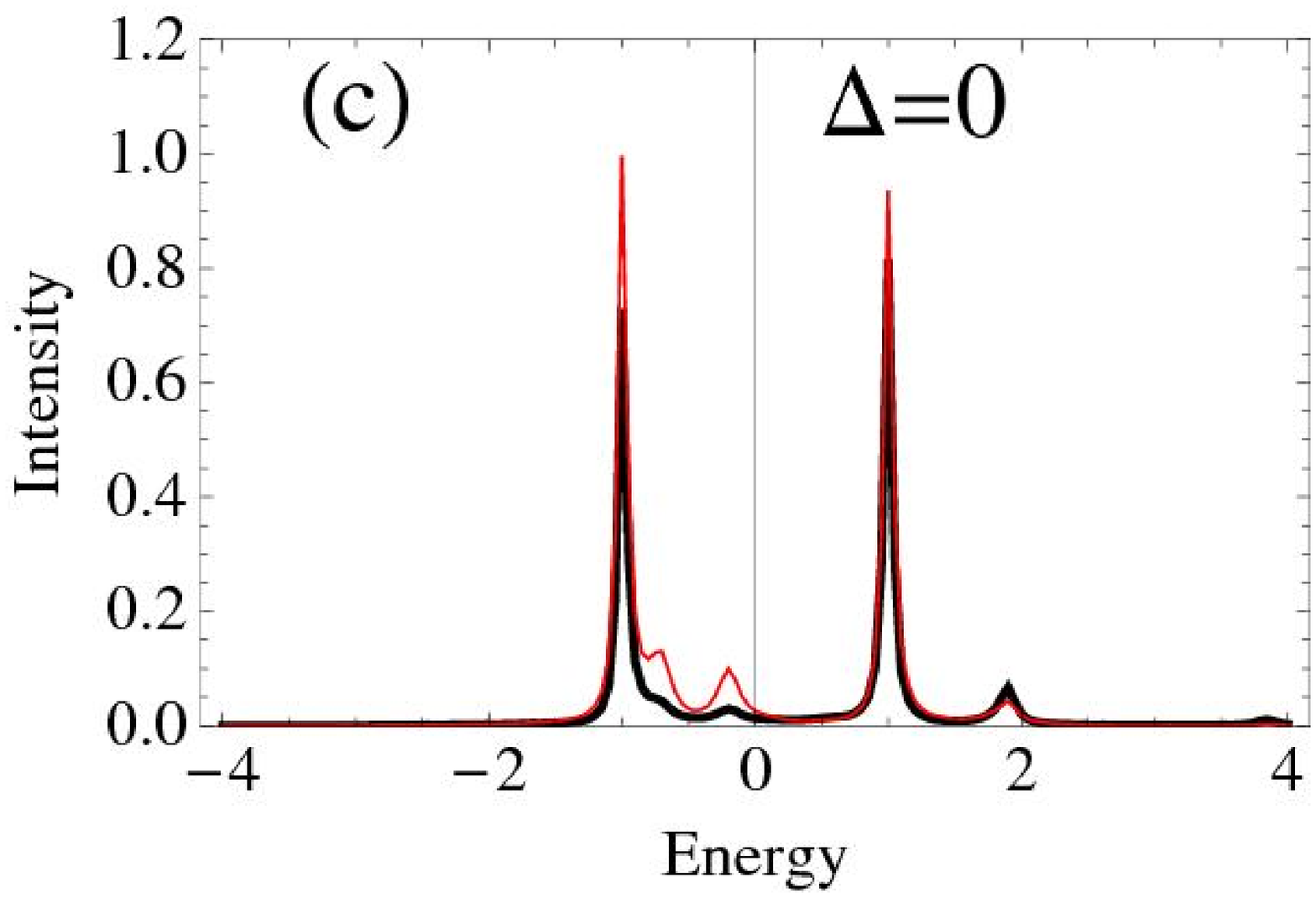, width=0.3\textwidth}
  \epsfig{file=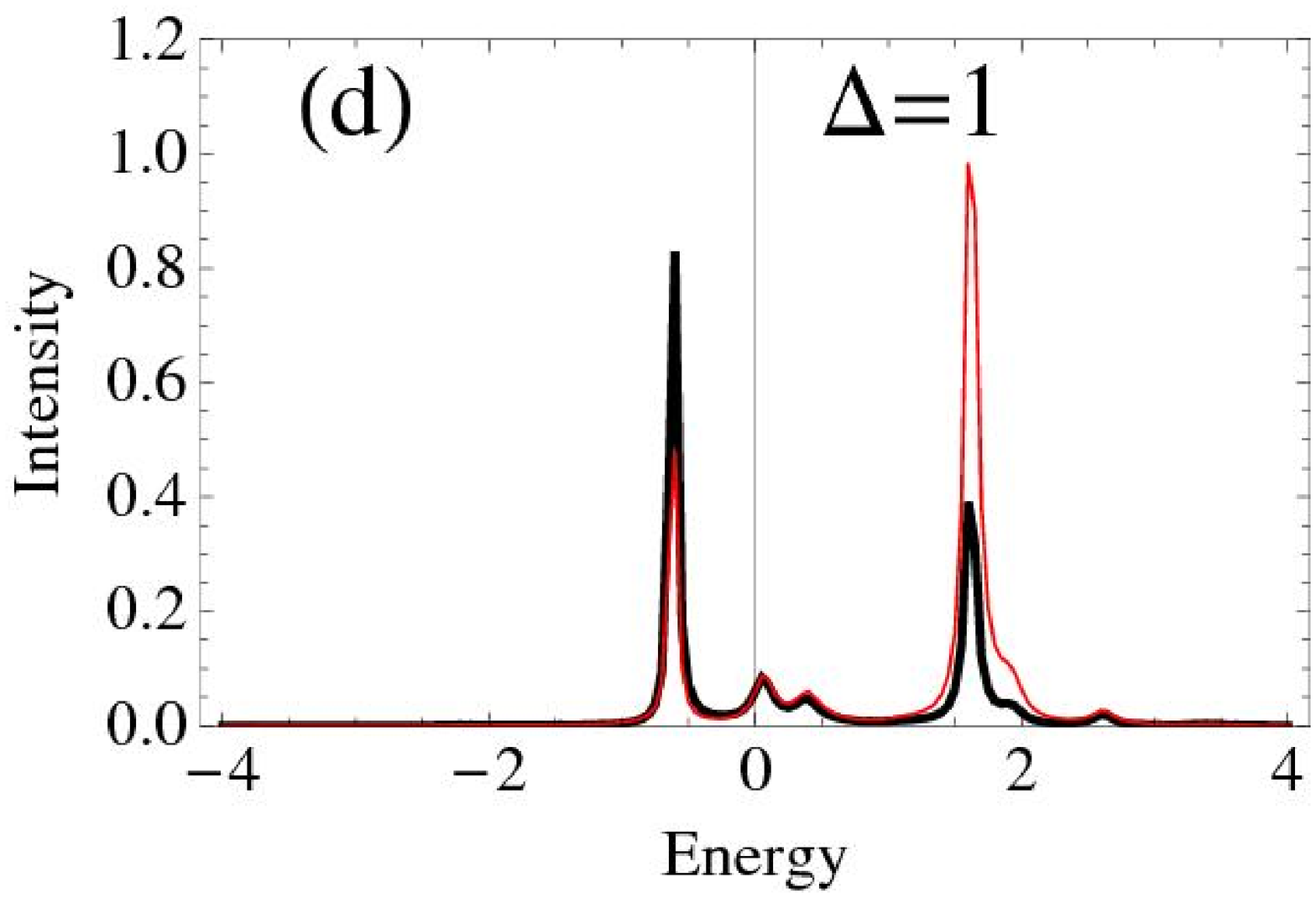, width=0.3\textwidth}
  \epsfig{file=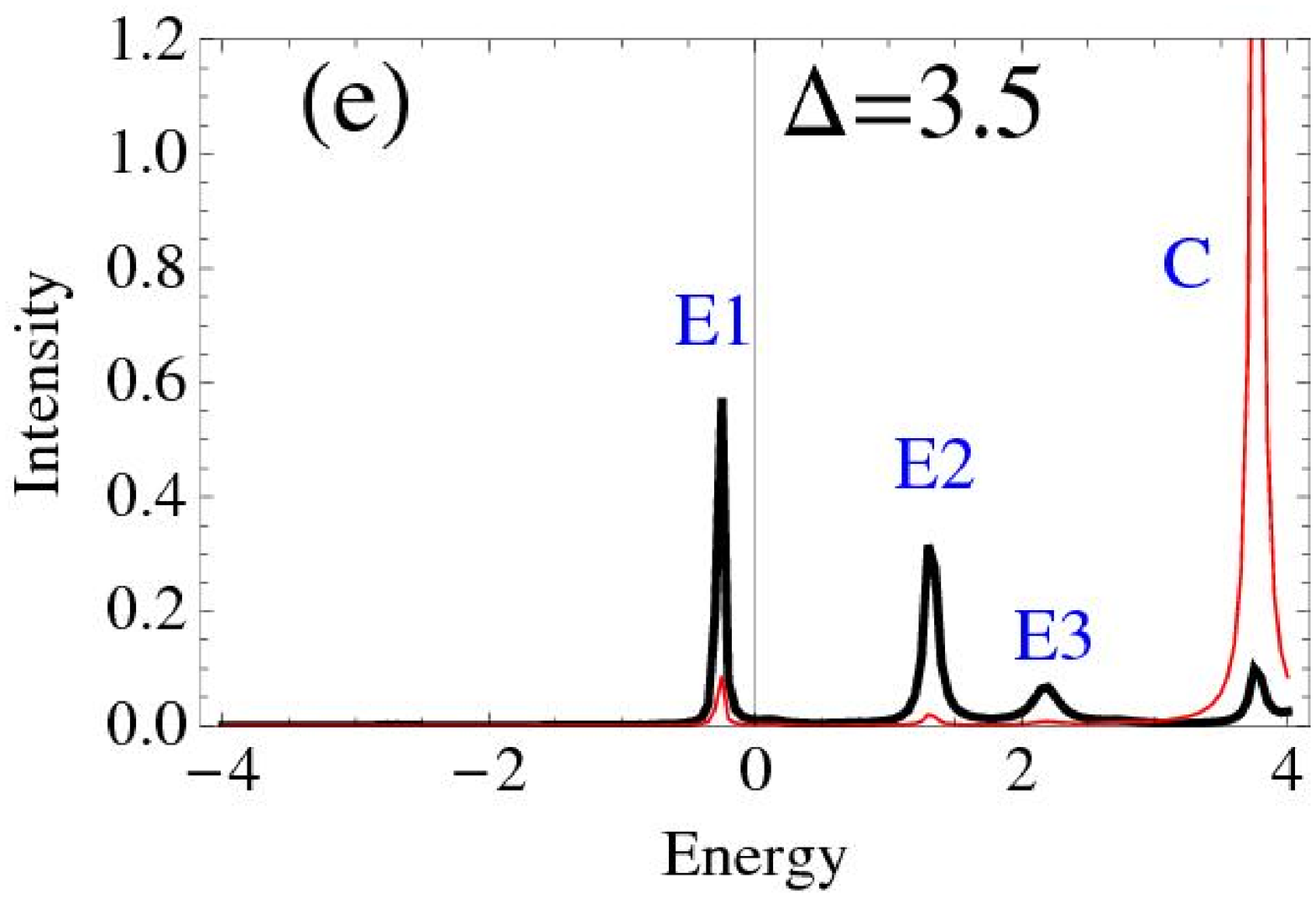, width=0.3\textwidth}
\end{center}
\caption{(Colour online) Spectra for different detunings corresponding
  to vertical ``cuts'' in Fig~(\ref{figure5}). Both cases, of cavity
  pumping ($P_1=P$, $P_2=0$) and electronic pumping ($P_1=0$,
  $P_2=P$), are represented in thin red and thick black respectively.
  At very large detunings ((a) and (e)), multiple excitons occupancy
  is observed through the peaks E1, E2, E3. Close to resonance, these
  result in satellites surrounding the linear Rabi doublet, that
  dominates because populations collapse at resonance.  Parameters:
  $\epsilon_2=0$, $U=2$, $\Gamma_1=0.1$, $\Gamma_2=0.01$, $P=0.01$.}
\label{figure6}
\end{figure}

Mean numbers of excitons and photons are plotted on
Fig.~(\ref{figure4}), for the two cases of cavity (only) and
electronic (only) pumping. Close to resonance, $\Delta\approx0$, both
pumping yield approximately equal exciton and photon populations (note
that the cavity pumping case has been magnified by a factor five).
Detuning the modes results in a collapse (cavity pumping) or increase
(electronic pumping) of the exciton population, as could be expected.
Regardless of the kind of pumping, however, the cavity population is
approximately constant. In the cavity pumping case, this is because
the exciton gets decoupled and thus the cavity is pumped at a constant
rate (one can actually see a small increase in its population). In the
electronic pumping case, this is because although the coupling
decreases, the exciton population increases in proportion so as to
feed the cavity with a constant flux of photons. In both cases, an
asymmetry is notable with detuning, because the interactions bring the
cavity and the exciton modes closer or further from resonance,
respectively, coupling them more efficiently for positive detuning and
therefore allowing a larger production of excitons in that case. As a
result, the nonlinear branches of the actual spectra (i.e., branches
others than the Rabi doublet) for positive and negative detunings,
shown on Fig.~(\ref{figure5}), are not exactly as those shown in
Fig.~(\ref{figure3})b. The blueshifted peak is more clearly seen in
the positive detuning case thanks to this exciton population asymmetry
with detuning. However, an excellent qualitative agreement is obtained
with the manifold method, if one superimpose the vacuum Rabi doublet
to the lines arising from higher manifolds.  Depending on the pumping
scheme---cavity (a) or exciton (b)---only quantitative features are
changed that consist mainly in different linewidths and intensities of
the branches, that are otherwise well accounted for by the manifold
method (see also Fig.~\ref{figure6}).  While comparing
Figs.~\ref{figure5}(a) and~(b), it should be borne in mind how the
total population changes with detuning, as shown on
Fig.~\ref{figure4}. For this reason, panel (b) has a more complex
structure, but this is due to the higher manifolds that can be reached
with the electronic pumping. It is in fact possible to identify the
third to second manifold contribution by extracting the lines in
Fig.~(\ref{figure5}) that do not appear in Fig.~(\ref{figure3})b.
These lines are clearly weaker due to the very low (but not vanishing)
probability to have three excitons in the system. The transitions from
even higher manifolds are too improbable to be seen in the spectra for
the pumping considered here. The main differences between the two
pumping schemes, if equal populations can be considered by adjusting
the pumping, are therefore to be found in the linewidth and
intensities of the lines. The positions of these lines embeds the most
precious indications on the physical system.

On Fig.~(\ref{figure6}), spectra are displayed for particular
detunings, in solid black for the case of electronic pumping, and thin
red for cavity pumping. The electronic pumping, which we repeat is the
most relevant case experimentally, yields the most interesting
spectral shape. On top of the Rabi doublet, the interactions produce
additional peaks that, at large detunings, are clearly associated to
the exciton (panels (a) and~(e)).  Three peaks, E1, E2 and~E3 are
obtained that correspond to one, two and three excitons coupling to
the cavity mode, respectively.  As these are brought in resonance with
the cavity mode, the linear Rabi doublet dominates (essentially
because the efficient coherent coupling collapses the exciton
population) and satellite peaks are observed, that betray the quantum
nature of the system, as the emission originates from transitions
between quantised manifolds.  In the absence of interactions, the Rabi
doublet is always observed independently of the total number of
particles (this is the definition of a \emph{linear}
behaviour). Therefore, interactions are useful to evidence a quantum
behaviour linked to quantized energy transfers, in the spirit of such
experiments as those used with atoms in cavity to demonstrate
quantization of the light field.\cite{brune96a} Here, nonlinear
features are observed directly in the optical spectrum, whereas in
Ref.~[\onlinecite{brune96a}], time-resolved measurements were used to
probe anharmonic oscillations of the Rabi flops. This represents a
notable experimental advantage, as measurements with cw incoherent
pumping are typically easier to perform than time-resolved
spectroscopy.

\begin{figure}[tbp]
  \begin{center}
    \epsfig{file=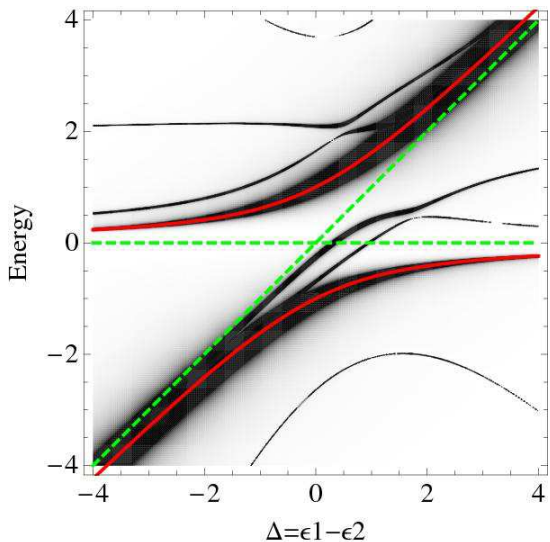, width=0.4\textwidth}
\end{center}
\caption{Cavity spectral emission as a funtion of detuning $\Delta$
  obtained via Green function technique. Parameters are similar to those in Fig.
  (\ref{figure5}).}
\label{figure7}
\end{figure}

We conclude this Section by discussing the computation of the emission
spectra in terms of nonequilibrium Green functions, that could be used
when the size of the Hilbert space forbids numerical computations with
a master equation. Keldysh Green functions are now routinely used in
description of electronic transport in mesoscopic systems\cite{hh96}
and we believe that their application for the description of optical
properties of QDs in the nonlinear regime is of methodological
interest. Besides, this approach allows the discrimination between
mean-field and correlation effects, which is important for the
understanding of the physical origin of the additional peaks in the
spectra of photoemission considered above. It can be shown (see
Appendix for details) that within a Hartree-Fock approximation the
emission spectra contains of a Rabi doublet and no satellite peaks
appear. This corresponds rather well to the case of the planar
microcavities with embedded quantum wells where the number of
excitations is large, but does not give an adequate description of the
spectra of QDs where the total number of excitations in the system is
small and correlation effects play a substantial role. Taking them
into account within the Keldysh Green function formalism allows to
reproduce the multiplet structure of the spectra, as shown in
Fig.~(\ref{figure7}) for the same parameters and occupation numbers as
in the case of cavity pumping in Fig.~(\ref{figure4}) and
(\ref{figure5}).  The Green function results agree with those obtained
with the density matrix formalism, the main difference being an
anticrossing between the satellite branches clearly seen at
Fig.~(\ref{figure7}), which is probably related to the truncation
scheme we use.

\section{Conclusions}
\label{conclusions}

We have analysed the optical spectra of a QD in strong coupling with
the single mode of a microcavity at the onset of the nonlinear regime
obtained through an incoherent pumping of either the cavity directly,
or of the exciton, with nonvanishing probabilities of more than one
excitation. We have considered a large quantum dot dominated by
exciton-exciton interactions. Exact dynamical results were obtained
with a density matrix master equation and the quantum regression
theorem. The Green function technique was also applied to support the
results and suggest possible extension to arbitrary high number of
excitations.

We showed that a deep physical understanding can be attained by
identifying the spectral lines with the transitions between quantised
energy levels of the nonlinear Hamiltonian. The spectral shapes
observed, that probe the first manifolds of multiple excitations, are
accounted mainly by the Coulomb energy shift on top of the linear Rabi
coupling, with crossings and anticrossings of the lines with detuning,
depending on their opposed (photon \emph{and} exciton) or identical
(photon \emph{or} exciton) character. The quantitative behaviour of
the linewidth and intensities of these lines depends mainly on the
dynamics of the system, such as the populations of the bare cavity
photons and bare excitons. Electronic pumping of the exciton---the
case most commonly realized experimentally as far as external
incoherent pumping is concerned---allows to achieve high populations
at nonzero detunings thanks to the small exciton lifetime and the
reduced effective coupling. The balance of these two factors
nevertheless retains an approximately constant cavity population. As a
consequence, the optimum experimental configuration to observe
nonlinear effects in the photoluminescence spectra is at an
intermediate, nonzero detuning, for instance when the detuning is
approximately equal to the coupling strength. There is an asymmetry
with the sign of the detuning due to the interactions that further
helps in characterising the nature of the nonlinearity (e.g., to which
extent it comes from the exciton-exciton Coulomb interaction).  The
presence of satellites with detuning demonstrates emission from
quantized manifolds, and as such is a signature of the quantum regime.
The spectral drift of these lines with detuning is a useful tool to
explicit the exact form of the Hamiltonian that accounts for the
exciton nonlinearities.

\section*{Acknowledgements}

We thank Prof. A.~V.~Kavokin, Prof. C.~Tejedor, Dr. R.~T.~Pepino and
Dr.~T.~C.~H.~Liew for useful discussions.  This work was supported by
Brazilian Ministry of Science and Technology, IBEM (Brazil) and by the
Spanish MEC under contracts Consolider-Ingenio2010 CSD2006-0019,
MAT2005-01388 and NAN2004-09109-C04-3. I. A. Shelykh acknowledges the
support of the grant of President of Russian Federation.

\appendix
\section{Optical spectra with the Keldysh Green function technique}

In this Appendix, we present the details of the calculations of the
photoemission spectrum with Keldysh Green function technique.

To model the pumping and decay of cavity photons and QD excitons
within the Keldysh Green function formalism, one needs to introduce
the Hamiltonian of the coupling of the photonic and excitonic fields
to the reservoirs of \emph{in}-coming and \emph{out}-going external
photons and excitonic reservoir. The latter corresponds to the pump of
the exciton state and its decay towards the leaky modes of the
system. The complete Hamiltonian can be thus represented as a sum of
that of eqn~(\ref{eq:H}) and the reservoir Hamiltonian
$H_\mathrm{ext}$ given by
\begin{multline}
  \label{eq:bath}
  H_\mathrm{ext}=
   \sum_{\mathbf{k}}(\omega_{2;\mathbf{k}}d_{2;\mathbf{k}}^\dagger d_{2;\mathbf{k}}
     + t_{2;\mathbf{k}} c_2 d_{2;\mathbf{k}}^\dagger+
  t_{2;\mathbf{k}}^* c_2^\dagger d_{2;\mathbf{k} })\\
{}+\sum_{\substack{\mathbf{k}\\{\eta=\mathrm{in,out}}}}
  (\omega_{1;\mathbf{k}} d_{1;\mathbf{k} \eta}^\dagger d_{1;\mathbf{k}
    \eta} + t_{1;\mathbf{k},\eta} c_1 d_{1;\mathbf{k}\eta}^\dagger+
  t_{1;\mathbf{k},\eta}^* c_1^\dagger d_{1;\mathbf{k} \eta})
\end{multline}

The operator $d_{1;\mathbf{k} \eta}$ ($d_{1;\mathbf{k} \eta}^\dagger$)
annihilates (creates) one incoming ($\eta=\mathrm{in}$) or outgoing
($\eta=\mathrm{out}$) free photon outside the cavity,
$d_{2;\mathbf{k}}$ annihilates a photon in a leaky mode, the
parameters $t_{1;\mathbf{k},\eta}$ and $t_{2;\mathbf{k}}$ are
cavity-reservoir and exciton-leaky mode coupling strengths
respectively and $\omega_{1,2;\mathbf{k}}$ are the dispersions of the
external photons and leaky modes. In the Keldysh formalism, the photon
spectral emission [counterpart of eqn~(\ref{Spec})] is recast as
\begin{equation}
  \label{Aph}
  S_{\mathrm{ph}}(\omega)=\frac{1}{\pi}\mathrm{Re}\>iG_{11}^<(\omega),
\end{equation}
where $G^<_{11}(\omega)$ is the lesser Green function in the frequency
domain for the photons inside the cavity.  In order to calculate the
correlator in eqn~(\ref{Aph}), we define the retarded Green function
$G_{jl}^{r}(t,t')=-i\theta(t-t_1) \langle [c_j (t); c_l^\dagger (t')]
\rangle$.\cite{hh96} The equation of motion of $G_{jl}^{r}(t,t')$
follows directly from the Heisenberg equation for the operators
$\dot{c}_{j}(t)=i [H+H_\mathrm{ext},c_j(t)]$:
\begin{eqnarray}
  \label{difeqGmunu}
  &(i \frac{\partial}{\partial t}-\epsilon_j)G_{jl}^{r}(t,t')=\delta(t-t')\delta_{jl}+V_R G_{\bar j l}^{r}(t,t')\nonumber\\
  & + \phantom{x} \delta_{j 2} U G_{jjj,l}^{r(2)}(t,t') + \int dt_1 \Sigma_{j l}^r(t,t_1) G_{j l}^{r}(t_1,t')\,,
\end{eqnarray}
where
\begin{eqnarray}
  \label{DefG2}
G_{jjj,l}^{r(2)}(t,t')=-i\theta(t-t_1) \langle [c_j^\dagger(t)
c_j(t) c_j (t); c_l^\dagger (t')]\rangle
\end{eqnarray}
and
\begin{eqnarray}
  \label{Defsigma}
  \Sigma_{jl}^r&=&\delta_{j 1} \sum_{\mathbf{k} \eta} |t_{1;\mathbf{k},\eta}|^2
  g_{1;\mathbf{k} \eta}^{r(0)}(t,t_1)\nonumber\\
  &+&\delta_{j 2} \sum_{\mathbf{k}}|t_{2;\mathbf{k}}|^2 g_{2;\mathbf{k}}^{r(0)}(t,t_1)\, ,
\end{eqnarray}
with $\overline{1}=2,\overline{2}=1$. The functions
\begin{eqnarray}
  \label{g0}
g_{1;\mathbf{k}\eta}^{r(0)}(t,t_1)&=&-i \theta(t-t_1)\langle[d_{1;\mathbf{k} \eta}(t) d_{1;\mathbf{k} \eta}^\dagger(t')]\rangle\nonumber\\
g_{2;\mathbf{k}}^{r(0)}(t,t_1)&=&-i \theta(t-t_1)\langle[d_{2;\mathbf{k}}(t) d_{2;\mathbf{k}}^\dagger(t')]\rangle
\end{eqnarray}
correspond to the retarded Green functions for free particles outside
the cavity (ingoing and outgoing photons, and leaky modes). To close
eqn~(\ref{difeqGmunu}), one needs the expression for
$G_{jjj,l}^{(2)}(t,t')$, which involves the third order Green
function.  The full procedure leads to an infinite set of equations,
coupling Green functions of different orders $G^{(n)}$ (in a so-called
\emph{Bogoliubov chain}). To close this system, one needs to apply a
truncation procedure, factorising Green functions of some given order
into products of lower-order correlators. The simplest way to close
the Bogoliubov chain is to apply the Hartree-Fock approximation, which
consists in breaking the correlator $G_{jjj,l}^{r(2)}(t,t')$ into
products of two-operator Green function.  More specifically, based on
the Wicks theorem, one can apply the following approximations:
\begin{eqnarray}
  \langle [c_{1}^\dagger (t) c_ 2(t) c_2(t),
  c_\nu^\dagger (t') ]\rangle &=& 2\langle c_{1}^\dagger c_ 2 \rangle
  \langle [c_2(t), c_\nu^\dagger (t') ]\rangle \nonumber \\ \langle
  [c_2^\dagger(t) c_2(t) c_1(t), c_\nu^\dagger (t')] \rangle &=& n_2
  \langle [ c_1(t), c_\nu^\dagger (t')] \rangle +\nonumber \\ &&
  \phantom{xxx}\langle c_2^\dagger c_1 \rangle \langle [c_2 (t) ,
  c_\nu^\dagger (t')] \rangle \nonumber \\ \langle [n_2(t) n_2(t)
  c_2(t), c_\nu^\dagger (t')] \rangle &=& 2 n_2 \langle [ n_2 (t)
  c_2(t), c_\nu^\dagger (t')] \rangle.\nonumber
\end{eqnarray}
However, in the problem considered, this mean-field approximation
appears to be too rough.  Indeed, in this case the resulting spectra
consist of a blue-shifted Rabi doublet without any additional
satellites (result not shown).

To take correlation effects into account, in the following, we keep
the correlators exact up to the second order. In this case, the
equation of motion for $G_{jmn,l}^{r(2)}(t,t')$ reads:
\begin{eqnarray}
  \label{eqmG2}
  &\left[i \frac{\partial}{\partial t}+\epsilon_j-\epsilon_n-\epsilon_m \right]G_{jmn,l}^{r(2)}(t,t')=\delta(t-t')b_{jmn,l}\nonumber\\
  &-V_R G_{\overline{j}mn,l}^{r(2)}(t,t')+V_R G_{j\overline{m}n,l}^{r(2)}(t,t')+V_R G_{jm\overline{n},l}^{r(2)}(t,t')\nonumber\\
  &+\delta_{n2}\delta_{mn} U G_{jnn,l}^{r(2)}(t,t')-\delta_{j2}U G_{jjjmn,l}^{r(3)}(t,t')\nonumber\\
  &+\delta_{m2}U G_{jmmmn,l}^{r(3)}(t,t')+\delta_{n2}UG_{jnmnn,l}^{r(3)}(t,t'),
\end{eqnarray}
where
\begin{equation}
  b_{jmn,l}(t)=\langle [c_j^\dagger (t) c_m(t) c_n(t), c_l^\dagger(t)] \rangle
\end{equation}
and
\begin{eqnarray}
  &G_{jmnpq,l}^{r(3)}(t,t')=-i\theta(t-t')\times\nonumber\\
  &\langle [c_j^\dagger(t) c_m^\dagger(t) c_n(t) c_p(t) c_q(t), c_l^\dagger(t')]\rangle \,.
\end{eqnarray}

According to the Wick's theorem for $G^{r(3)}$, the following
truncation procedure holds:
\begin{eqnarray}
  \label{trunc}
 &&G_{j m n p q}^{r(3)}(t,t') = - i \theta(t-t') \times \nonumber \\ && \{
\langle c_j^\dagger (t) c_n(t)\rangle \langle [ c_m^\dagger (t) c_p(t) c_q(t), c_l^\dagger(t') ] \rangle + \nonumber \\
&& \langle c_j^\dagger (t) c_p(t)\rangle \langle [ c_m^\dagger (t) c_n(t) c_q(t), c_l^\dagger(t') ] \rangle + \nonumber \\
&& \langle c_j^\dagger (t) c_q(t)\rangle \langle [ c_m^\dagger (t) c_n(t) c_p(t), c_l^\dagger(t') ] \rangle + \nonumber \\
&& \langle c_m^\dagger (t) c_n(t)\rangle \langle [ c_j^\dagger (t) c_p(t) c_q(t), c_l^\dagger(t') ] \rangle + \nonumber \\
&& \langle c_m^\dagger (t) c_p(t)\rangle \langle [ c_j^\dagger (t) c_n(t) c_q(t), c_l^\dagger(t') ] \rangle + \nonumber \\
&& \langle c_m^\dagger (t) c_q(t)\rangle \langle [ c_j^\dagger (t) c_n(t) c_p(t), c_l^\dagger(t') ] \rangle \}.
\end{eqnarray}

Substituting eqn~(\ref{trunc}) in eqn~(\ref{eqmG2}) we obtain a new
equation for $G^{r(2)}$ only,
\begin{eqnarray}
  \label{G2big}
  &&\left[i \frac{\partial}{\partial t}+\epsilon_j-\epsilon_n-\epsilon_m \right]G_{jmn,l}^{r(2)}(t,t')=\delta(t-t')b_{jmn,l}\nonumber\\
  &&-V_R G_{\overline{j}mn,l}^{r(2)}(t,t')+V_R G_{j\overline{m}n,l}^{r(2)}(t,t')+V_R G_{jm\overline{n},l}^{r(2)}(t,t')\nonumber\\
  &&+\delta_{n2}\delta_{mn} U G_{jnn,l}^{r(2)}(t,t')-\delta_{j2} U \times \nonumber \\ && \{ 2 n_{jj} G_{jmn,l}^{r(2)}(t,t') + 2 n_{jm} G_{jjn,l}^{r(2)}(t,t') + 2 n_{jn} G_{jjm}^{r(2)}(tt')\}\nonumber\\
  &&+\delta_{m2}U \{2 n_{jm}G_{mmn.l}^{r(2)}(t,t')+n_{jn}G_{mmm,l}^{r(2)}(t,t')\nonumber\\ && \phantom{xxx} + 2 n_{mm} G_{jmn,l}^{r(2)}(t,t')+n_{mn} G_{jmm,l}^{r(2)}(t,t')\}\nonumber\\&&+
  \delta_{n2}U \{ n_{jm} G_{nnn,l}^{r(2)} (t,t') + 2 n_{jn} G_{nmn,l}^{r(2)}(t,t') \nonumber \\ && \phantom{xxx} + n_{nm} G_{jnn,l}^{r(2)}(t,t') +2 n_{nn} G_{jmn,l}^{r(2)}(t,t')\}\, ,
\end{eqnarray}
with $n_{jl}=\langle c_j^\dagger c_l \rangle$. Defining the Green
function vector
\begin{equation}
  \underline{\mathbf{G}}^{r(2)}_{\phantom{.}l}=
  (G_{222,l}^{r(2)},
  G_{122,l}^{r(2)},
  G_{212,l}^{r(2)},
  G_{121,l}^{r(2)},
  G_{111,l}^{r(2)},
  G_{211,l}^{r(2)})^T
\end{equation}
($T$ is for transpose), eqn~(\ref{G2big}) can be written in a matrix
form
\begin{equation}
  \label{eqmmatrixG2_2}
  \underline{\mathbf{G}}^{r(2)}_{\phantom{.}l} (t,t')= \underline{\underline{\mathbf{G}}}^{r0}(t,t') \underline{\mathbf{b}}_{\phantom{.}l} + \int dt_1 \underline{\underline{\mathbf{G}}}^{r0}(t,t_1) \underline{\underline{\mathbf{M}}}\phantom{x}\underline{\mathbf{G}}^{r(2)}_{\phantom{.}l} (t_1,t'),
\end{equation}
where the vectors are underlined once and matrices twice. The matrix
of zero order Green functions
$\underline{\underline{\mathbf{G}}}^{r(0)}$ is given by
\begin{equation}
  \label{G0rmatrix}
  \underline{\underline{\mathbf{G}}}^{r(0)}(t,t')=
\mathrm{Diag}(g_{2}^{r(0)},g_{2}^{r(0)},g_{2}^{r(0)},g_{1}^{r(0)},g_{1}^{r(0)},g_{1}^{r(0)})
\end{equation}
where~$\mathrm{Diag}(\mathbf{r})$ is the matrix with elements
of~$\mathbf{r}$ on its diagonal, zero elsewhere, and $g_{n}^{r(0)}$ is
defined according to $\left[i \frac{\partial}{\partial
    t}-\epsilon_n\right]g_{n}^{r(0)}(t,t')=\delta (t-t')$. The matrix
$\underline{\underline{\mathbf{M}}}$ is the sum of two terms,
$\underline{\underline{\mathbf{M}}}=\underline{\underline{\mathbf{M}}}_R
+ \underline{\underline{\mathbf{M}}}_U$, where
\begin{equation}
  \underline{\underline{\mathbf{M}}}_R=\left(
    \begin{array}{cccccc}
      0 & -V_R & 2V_R & 0 & 0 & 0 \\
      -V_R & -\Delta & 0 & 2V_R & 0 & 0\\
      V_R & 0 & \Delta & -V_R & 0 & V_R \\
      0 & V_R & -V_R & -\Delta & V_R & 0 \\
      0 & 0 & 0 & 2V_R & 0 & -V_R \\
      0 & 0 & 2V_R & 0 & -V_R & \Delta \\
\end{array}
\right),
\end{equation}
and
\begin{equation}
  \underline{\underline{\mathbf{M}}}_U = U\left(
    \begin{array}{cccccc}
      1+6 n_{22} & 0 & 0 & 0 & 0 & 0 \\
      6 n_{12} & 1+6 n_{22} & 0 & 0 & 0 & 0\\
      0 & 0 & 0 & 0 & 0 & 0 \\
      n_{11} & n_{21} & 2 n_{12} & 2 n_{22} & 0 & 0 \\
      0 & 0 & 0 & 0 & 0 & 0 \\
      0 & 0 & -4 n_{21} & 0 & 0 & -2 n_{22} \\
\end{array}
\right).
\end{equation}
Finally, the vector $\underline{\mathbf{b}}_{\phantom{.}l}$ is defined
as
\begin{equation}
  \underline{\mathbf{b}}_{\phantom{.}l}=\left(
\begin{array}{cccccc}
  2\delta_{2l}n_{22} \\
  2\delta_{2l} n_{12} \\
  \delta_{2l} n_{21} + \delta_{1l} n_{22}  \\
  \delta_{2l}  n_{11} + \delta_{1l} n_{12} \\
  2\delta_{1l} n_{11} \\
  2\delta_{1l} n_{21} \\
\end{array}
\right).
\end{equation}

The quantities $n_{jl}$ can be computed self-consistently together
with the lesser Green function through the expression $n_{jl}=(i/2\pi)
\int G_{lj}^<(\omega)d\omega$. For simplicity, here we take $n_{jl}$
as numerical parameters. That is, we substitute the effect of the
coupling to external reservoirs for an effective decay added as an
imaginary part to the bare cavity energy $\epsilon_1-i\Gamma_1/2$ and
to the bare exciton energy $\epsilon_2-i \Gamma_2/2$. The effect of
the pump is partially included by adjusting $n_{11}$ and $n_{22}$ to
the values calculated in Fig.~(\ref{figure4}).  It is possible to show
that the Keldysh-contour-ordered Green function
$G_{jmn,l}^{(2)}(\tau,\tau')=-i \langle [T_C c_j^\dagger(\tau)
c_m(\tau) c_n(\tau), c_{l}^\dagger(\tau')]\rangle$ has an equation
formally similar to eqn~(\ref{eqmmatrixG2_2}).\cite{hh96} Then,
applying analytical continuation rules, we find in the frequency
domain the equations for the retarded ($r$) and the advanced ($a$)
Green functions as well as for the lesser ($<$) Green function,
\begin{equation}
\label{G2r}
  \underline{\mathbf{G}}^{r,a(2)}_{\phantom{.}l} (\omega)= \underline{\underline{\mathbf{G}}}^{r,a (0)}(\omega) \underline{\mathbf{b}}_{\phantom{.}l} +  \underline{\underline{\mathbf{G}}}^{r,a (0)}(\omega) \underline{\underline{\mathbf{M}}}\phantom{x}\underline{\mathbf{G}}^{r,a (2)}_{\phantom{.}l} (\omega),
\end{equation}
and
\begin{eqnarray}
\label{G2lesser}
\underline{\mathbf{G}}^{< (2)}_{\phantom{.}l} (\omega)&=& \underline{\underline{\mathbf{G}}}^{< (0)}(\omega) \underline{\mathbf{b}}_{\phantom{.}l} +\underline{\underline{\mathbf{G}}}^{r (0)}(\omega) \underline{\underline{\mathbf{M}}}\phantom{x}\underline{\mathbf{G}}^{< (2)}_{\phantom{.}l} (\omega)\nonumber\\
&+&\underline{\underline{\mathbf{G}}}^{< (0)}(\omega) \underline{\underline{\mathbf{M}}}\phantom{x}\underline{\mathbf{G}}^{a (2)}_{\phantom{.}l} (\omega).
\end{eqnarray}

In order to take into account the finite lifetimes of the exciton and
photon modes, a phenomenological broadening was included in the lesser
Green function of the free particles, which reads:
\begin{equation}
  [\underline{\underline{\mathbf{G}}}^{< (0)}(\omega)]_{j}=-2i \frac{\frac{\Gamma_j}{2}}{(\omega-\epsilon_j)^2+(\frac{\Gamma_j}{2})^2}n_{jj}\,,
\end{equation}
where $j=2$ for the first three and $j=1$ for the last three diagonal
elements of the matrix $\underline{\underline{\mathbf{G}}}^{<
  (0)}(\omega)$.

Finally, we note that the equation for the retarded Green function of
the first order can be written in a matrix form in the frequency
domain,
\begin{eqnarray}
\label{eqmmatrixGr}
&  \left(
\begin{array}{cc}
  \omega-\epsilon_1+i\frac{\Gamma_1}{2} & -V_R \\
  -V_R & \omega-\epsilon_2+i\frac{\Gamma_2}{2} \\
\end{array}
\right)\left(
\begin{array}{cc}
  G_{11}^r & G_{12}^r \\
  G_{21}^r & G_{22}^r \\
\end{array}
\right)= \nonumber\\
&
\left(
  \begin{array}{cc}
  1 & 0 \\
  0 & 1 \\
\end{array}
\right)
+U \left(
  \begin{array}{cc}
 0 & 0 \\
 G_{222,1}^{r(2)} & G_{222,2}^{r(2)} \\
\end{array}
\right)\,,
\end{eqnarray}
while for the lesser Green function of the first order one has
\begin{eqnarray}
\label{eqmmatrixGlesser}
 \left(
   \begin{array}{cc}
     \omega-\epsilon_1+i\frac{\Gamma_1}{2} & -V_R \\
     -V_R & \omega-\epsilon_2+i\frac{\Gamma_2}{2} \\
\end{array}
\right)\left(
\begin{array}{cc}
  G_{11}^< & G_{12}^< \\
  G_{21}^< & G_{22}^< \\
\end{array}
\right)= \nonumber\\
\left(
\begin{array}{cc}
  \Sigma_{11}^< & 0 \\
  0 & 0 \\
\end{array}
\right)\left(
\begin{array}{cc}
  G_{11}^a & G_{12}^a \\
  G_{21}^a & G_{22}^a \\
\end{array}
\right)
+ U \left(
\begin{array}{cc}
  0 & 0 \\
  G_{222,1}^{<(2)} & G_{222,2}^{<(2)} \\
\end{array}
\right)\,.
\end{eqnarray}

Solving the set of eqns~(\ref{G2r})-(\ref{eqmmatrixGlesser}) we
obtain the emission spectrum of the system, cf.~eqn~(\ref{Aph}).

\bibliography{Sci,books,references}

\end{document}